\documentstyle[aps,12pt]{revtex}

\begin{document}

\title{Winding Angle Distributions for Directed Polymers}
\author{Barbara Drossel and Mehran Kardar}
\address{Department of Physics, Massachusetts Institute of
Technology, Cambridge, Massachusetts 02139}
\date{\today}
\maketitle
\begin{abstract}
We study analytically and numerically the winding of directed polymers of lenght $t$ around each other or around a rod. Unconfined polymers in pure media have exponentially decaying winding angle distributions, the decay constant depending on whether the interaction is repulsive or neutral, but not on microscopic details. In the presence of a chiral asymmetry, the exponential tails become non universal. In all these cases the mean winding angle is proportional to $\ln t$. When the polymer is confined to a finite region around the winding center, e.g. due to an attractive interaction, the winding angle distribution is Gaussian, with a variance proportional to $t$. 
We also examine the windings of polymers in random systems. Our results suggest that randomness reduces entanglements, leading to a narrow (Gaussian)
distribution with a mean winding angle of the order of $\sqrt{\ln t}$. 
\end{abstract}
\section{Introduction and Summary}
\label{intro}
The topological constraints produced by the windings of polymers\cite{Gen71}
strongly affect the dynamics of polymer solutions. As a consequence of polymer
entanglement, the viscosity of a solution of polymers above the overlap
concentration is many orders of magnitude higher than the viscosity of the
solvent. An analytical treatment of these topological constraints is extremely
difficult, and  theoretical efforts therefore focus on the limit of high
polymer concentrations, where effective medium theories and the tube model
successfully describe several aspects of the dynamics of the polymer solution
\cite{doi86}, or on the limit of only one or two polymers, where the different
possible configurations can be studied explicitely \cite{gro93}.

In this article, we take the latter approach, focussing on the winding of a
directed polymer (DP) around a rod, or of two DPs around each other, as shown
in Fig.~\ref{picture}. DPs have a preferred direction $\hat \tau$, and their
configuration can be described by the function $\{\vec r(\tau)\}$, with $\tau
\in [0,t]$, where $\vec r = (x_1,x_2)$ is the coordinate in the plane
perpendicular to the preferred direction. Going to relative coordinates $ \vec
r_2(\tau) - \vec r_1(\tau)$, the winding of two DPs around each other can be
mapped to the winding of a single DP around a rod (see section \ref{free}).
Since DPs cannot have knots, their main topological constraints are windings.

Although less common than flexible polymers, DPs are a good model of several
semiflexible and rigid polymers.  Example are biological macromolecules such as
DNA or liquid crystals composed of stacks of disk-shaped molecules.
These polymers are aligned parallel to each other when their concentration is
sufficiently large, forming crystalline and liquid crystalline phases (for a
review on statistical mechanics of DPs see e.g. \cite{nel94}).  Isolated DPs
can be realized by embedding a long polymer in a nematic solvent
\cite{deGen82}. Another important class of directed ``polymers'' are magnetic
flux lines in high-$T_c$ superconductors that are oriented parallel to the
direction of the external magnetic field. Due to the high temperature in the
system and the weak coupling between different layers in the superconductor,
thermal fluctuations of the flux lines are considerable, leading to
entanglements \cite{nel88,obu90} and windings around  columnar pins
\cite{nel93}.

In order to calculate the winding angle distribution of DPs, we map them onto
two-dimensional walks $\{\vec r(\tau)\}$, where the arc length $\tau$ plays the
role of the time coordinate. The winding angle distribution depends on the
interaction between the polymer and the winding center, and the properties of
the embedding medium. In the following three sections, we discuss three
different classes of winding angle distributions, for each of which the scaling
variable is a different combination of the winding angle $\theta$ and the
polymer length $t$. In section \ref{free}, we consider DPs in an infinitely
large pure medium. These polymers can be mapped on ideal random walks, the mean
horizontal distance $\langle |\vec r(\tau) - \vec r(0)|\rangle $ from the
starting point increasing with the square root of $\tau$. The number of returns
to the winding center is proportional to $\ln t$ for such a random walk. We
will see below that because of the finite return probability to the winding
center even for large times,  the winding angle distribution depends on
properties of the winding center. We will find different winding angle
distributions depending on whether the interaction between the winding center
and the polymer is a hard-core repulsion or is neutral, and whether the winding
center shows a chiral asymmetry. However, the winding angle distribution does
not depend on microscopic details like the shape of the winding center or a
possible underlying lattice structure. The probability distribution for the
winding angle depends in the limit $t \to \infty$ only on the combination $x =
2\theta / \ln(t)$ of the winding angle and the length of the walk, and not on
each variable separately. For large $|x|$, all three mentioned winding angle
distributions decay exponentially in $|x|$. The scaling variable proportional
to  $\theta / \ln(t)$ can be explained as follows:
After time $t$, the walker has a typical distance $r(t) \propto \sqrt{t}$ from
the
starting point, which is chosen to be close to the winding center. Assuming
that $r(t)$ is the only relevant length scale, dimensional arguments, combined
with the Markovian property, suggest that $dr/d\theta=r f(\theta)$. The
rotational invariance of the system implies that $f(\theta)$ must be a
constant, i.e.  the increase in winding angle cannot depend on the number of
windings or angular  position. Hence,
\begin{equation}\label{idealRW}
d\theta \propto {dr \over r}\propto {dt \over t} = d(\ln t)\, ,
\end{equation}
leading to a scaling variable proportional to $\theta / \ln t$.

If the interaction between the polymer and the winding center is attractive,
the polymer can be bound to the winding center, and its transverse wandering is
limited.  Polymers can also be confined by a finite container or by neighboring
polymers. In all these cases, polymer segements of length $\Delta t$ that are
small compared to the total length $t$, but large compared to the
length needed to make a winding, have identical winding angle distributions.
Applying the central limit theorem, we conclude that the total winding angle
distribution is a Gaussian with a scaling variable $\theta^2 /t$. This
situation will be discussed in section \ref{confine}.

Finally, we discuss in section \ref{nonideal} certain DPs that cannot be
described by ideal random walks, since they do not satisfy a Markov property.
Polymers that are embedded in a random medium, e.g. a gel, have an energy that
depends on the polymer configuration. Similarly, the energy of magnetic flux
lines in high-$T_c$ superconductors with  point defects depends on their
configuration. The mean distance from the starting point for these polymers in
random media increases faster than for ideal walks, since the line searches for
low-energy configurations. The number of returns to the winding center remains
 finite in the limit of infinite length. Consequently, the interaction of the
line with the winding center does not affect the winding angle distribution, as
long as it is not strong enough to bind the polymer. We will see that the
winding angle distribution of polymers in random media is a Gaussian with a
variance proportional to $ \ln t$, leading to a  scaling variable $x
\propto\theta/\sqrt{\ln(t)}$. This means that the pinning to randomness
decreases the mean winding angle from the order of $\ln t$ to the order of
$\sqrt{\ln t}$. Interestingly, this winding angle distribution is similar to
the one for two-dimensional self-avoiding random walks \cite{dup88}. The
following scaling argument explains why the winding angle distribution in both
situations is Gaussian with a variance proportional to $\ln t$:
Starting from the origin divide the walk into
segments of 1, 2, $\cdots,\, 2^n\approx t/2$ steps. Since the $\alpha^{\rm th}$
segment is at a distance of roughly $2^{\alpha\nu}$ from the center (with
$\nu = 3/4$) and has a characteristic size of the same order, it is reasonable
to assume that each segment spans a random angle $\theta_\alpha$ of order one.
Under the mild assumption that the sum $\theta=\sum_{\alpha=1}^n \theta_\alpha$
satisfies the central limit theorem, we then conclude that $\theta$ is Gaussian
distributed with a variance proportional to $n\propto\ln t$. Since this
argument relies on the irrelevance of the winding center, it cannot be applied
to the distributions in section \ref{free}.

Many results of this article have been reported previously in
Ref.~\cite{dro96}. They point out the rich behavior already present in
the simplest of problems involving topological defects. Properties of the
winding center, interactions, various
types of randomness are all potentially relevant, leading to different
universal distribution functions. The concluding section  \ref{concl} of this
article gives an outlook on possible further universality classes and on the
winding of non-direceted polymers.

\section{Winding angle distributions in an infinite homogeneous medium}
\label{free}

In this section, we study winding angle distributions of DPs in infinite
homogeneous media, all characterized by a scaling variable $x = 2\theta / \ln
t$ and exponential tails. We keep the initial point of the polymers fixed, but
otherwise allow them to move freely. The precise form of the winding angle
distribution depends on the interaction with the winding center. In  subsection
\ref{absorbing}, we consider two DPs with hard-core repulsion, or,
equivalently, one DP winding around a repulsive rod, leading to the
distribution in Eq.~(\ref{abs}). For neutral winding centers, the corresponding
winding angle distribution given in Eq.~(\ref{refl}) has a decay constant that
is smaller by one half (subsection \ref{neutral}). These two distributions
occur under fairly general conditions (see subsection \ref{univers}). However,
when the symmetry with respect to the sign of the winding angle  is broken,
 new (asymmetric) distributions occur (subsection \ref{chiral}), with the decay
constants of the exponential tails depending on the degree of chirality.

\subsection{Winding in the presence of hard-core repulsion}
\label{absorbing}

\subsubsection{Mapping to a random walk with absorbing boundary conditions}

The energy of a given configuration of two DPs $\vec r_1(\tau)$ and $\vec
r_2(\tau)$ of length $t$ is given by
\begin{equation}
E\left[\vec r_1(\tau), \vec r_2(\tau)\right] = \int_0^t d\tau \left[
c\left(d\vec r_1\over d \tau\right)^2 + c\left(d\vec r_2\over d \tau\right)^2 +
V(\vec r_1 - \vec r_2)\right]  .\label{energy}
\end {equation}
The potential $V(\vec r)$ has a hard-core, $V(r) = \infty$ for $r < a$, and
$V(r) = 0$ for $ r > a $. The first two terms are the elastic energies of the
polymers, where the parameter $c$ is related to their stiffness.
Introducing the relative coordinate $\vec r = \vec r_1 - \vec r_2$ and the
center-of-mass coordinate $\vec R = (\vec r_1 + \vec r_2) / 2$,
Eq.~(\ref{energy}) becomes
$$ E\left[\vec R(\tau), \vec r(\tau)\right] = \int_0^t d\tau \left[ {c\over
2}\left(d\vec r\over d \tau\right)^2 + 2c\left(d\vec R\over d \tau\right)^2 +
V(\vec r)\right] .
$$
The partition function for the two polymers is
\begin{equation}
Z = \int{\cal D}\left[\vec R(\tau)\right]{\cal D} \left[\vec r(\tau)\right]
\exp\left\{-E\left[\vec R(\tau), \vec r(\tau)\right]  /  k_B T\right\},
\end{equation}
where the integral is taken over all possible configurations $[\vec R(\tau)]$
and $[\vec r(\tau)] $.  The expression ${\cal D} [\vec r(\tau)]$ denotes a path
integral and is the continuum limit of $\prod_{i = 1}^n (\int d\vec
r(\tau_i))$,
$k_B$ is the Boltzmann constant, and $T$ is the temperature.

As long as we are only interested in quantities related to the relative
coordinate, like the winding angle, we can integrate out the center-of-mass
variations, and focus on the partition function for the relative coordinate
alone, i.e.,
\begin{equation}
Z = \int{\cal D} \left[\vec r(\tau)\right] \exp\left\{-\int_0^t d\tau \left[
{c\over 2}\left(d\vec r\over d \tau\right)^2 +  V(\vec r)\right] / k_B
T\right\}. \label{partfunc}
\end{equation}
This is identical to the partition function for a single DP winding around a
rod. Due to the hard-core repulsion, all configurations where the polymer and
the rod penetrate each other, do not contribute to the partition function ($V =
\infty$), while  $V = 0$ for all other configurations.

A two-dimensional random walk can be described by the Langevin-equation
\begin{equation}
{d \vec r \over dt} = \vec \eta(t), \label{rw}
\end{equation}
where $\eta$ is a stochastic force with zero mean ($\langle \eta(t) \rangle =
0$) and the correlation function
$\langle \vec \eta(t) \vec \eta(t')\rangle = 2 D \delta(t - t').$
The probability distribution of $\vec \eta$ is a Gaussian,
i.e.
$$P\left[\eta(t)\right] \propto \exp\left[-D \left(\vec
\eta(t)\right)^2\right]\, .$$
With Eq.~(\ref{rw}), we find that the probability for a given trajectory $[\vec
r(\tau)]$ of the random walk is proportional to
$$
\exp\left\{-\int_0^t d\tau \left[ {D\left(d\vec r\over d \tau\right)^2 }\right]
\right\}.
$$
When  all walks that enter a region of radius $a$ around the origin get
absorbed, the probability that the random walk has a trajectory $[\vec
r(\tau)]$ is identical to the probability that the above DP has the
cofiguration $[\vec r(\tau)]$ (compare to Eq.~(\ref{partfunc}), with $D = c/k_B
T$).
This correspondence between DPs with hard-core repulsion and random walks with
absorbing boundary conditions was first pointed out by Rudnick and Hu
\cite{rud87}.

\subsubsection{Conformal mapping of the random walk}

Since we are interested in the winding angle of the random walk, it is
convenient to perform a transformation such that the winding angle becomes one
of the coordinates. To this purpose,
we represent the walk $\vec r(t) = \left(x_1(t), x_2(t)\right)$ by the complex
number
$$
z(t)=x_1(t)+ix_2(t)\, .
$$
The time evolution of each random walker satisfies
\begin{equation}
dz=\eta(t) dt,
\end{equation}
where $\eta(t)$ is now complex, with
\begin{equation}
\left\langle \eta(t)\eta^*(t') \right\rangle=2D\delta(t-t').
\end{equation}
We now introduce the new variable $\zeta$ by the transformation
\begin{equation}
\zeta(t)=\ln z(t)=\ln r(t) +i\theta(t), \label{complvar}
\end{equation}
where $\rho=\ln r =\ln\sqrt{x_1^2+x_2^2}$.
Since $d\zeta=\eta(t)dt/z(t)$, the stochastic motion of the walker in the
new complex plane is highly correlated to its location, i.e., the walk is no
longer random.
This feature can
be removed by defining a new time variable
\begin{equation}
d{\tilde t}={dt\over |z(t)|^2}
\end{equation}
{\it for each walker}, which leads to
\begin{equation}
d\zeta=\mu({\tilde t})d{\tilde t}, \quad{\rm with}\quad\mu({\tilde
t})=z^*(t)\eta(t).
\end{equation}
Since
\begin{equation}
\left\langle \mu({\tilde t})\mu^*({\tilde t}')
\right\rangle=2D|z(t)|^2\delta(t-t')
=2D\delta({\tilde t}-{\tilde t}'),
\end{equation}
the evolution
of $\zeta({\tilde t})$ is that of a random walk. Under the transformation in
Eq.~(\ref{complvar}), the absorbing disc in the plane $z$ maps onto an
absorbing wall in the plane $\zeta$ (see Fig.~\ref{trafo}).

For simplicity we choose the initial condition $\zeta(t={\tilde t}=0)=0$, i.e.
the original walker starts out at $z=1$. We also set the diffusion constant
to $D = 1/2$, so that the mean square distance over which the walk moves
during a time $t$ is $\langle r^2(t)\rangle = t$. Consequently,
the probability that
$r(t)$ is within an
interval $[\sqrt{\pi} t^{(1 - \epsilon)/2}, \sqrt{\pi} t^{(1 + \epsilon)/2}]$
around its mean value of $\sqrt{\pi} t$, is
\begin{eqnarray}
p(t,\epsilon) &=& \int_{\sqrt{\pi} t^{(1 - \epsilon)/2}}^{\sqrt{\pi}
t^{(1 + \epsilon)/2}} {\exp\left(-r^2/2t\right)\over 2\pi t} 2\pi r\, dr
\nonumber \\
&=& \int_{\pi t^{-\epsilon}/2}^{\pi t^\epsilon /2} \exp(-s) ds,
\label{eqepsilon}
\end{eqnarray}
and approaches unity in the limit $t \to \infty$. The effect of the absorbing
disc on this probability can be neglected in the limit $t \to \infty$, since
the disc becomes smaller when viewed from larger distances. In this limit, the
distance
$r$ from the starting point $z=1$ is identical to the distance from the
origin, and  $p(t,\epsilon)$ is identical to the probability that
$\zeta({\tilde t})$ is in the interval
$[0.5(1 - \epsilon)\ln t,0.5(1 +\epsilon)\ln t]$. So the endpoints of all walks
(except for an infenitesimal fraction) that take a time $t$ in the original
plane map within a strip of width $\epsilon \ln t$ in the $\zeta$-plane, as
indicated in Fig.~\ref{trafo}.
If we shrink the complex plane $\zeta$ by a factor of $(\ln t)/2$,
the walker is within a distance $\epsilon$ of
the line with real value of unity. Thus, all walks of length $t$ in the
$z$-plane are mapped on walks that end at the line with real value of unity,
without having gone beyond (see Fig.~\ref{trafo}).
Since there is a separate
transformation ${\tilde t}(t)$ for each walker, walks of the same
length $t$ map on walks of different length ${\tilde t}$. (To be precise,
we also
have to shrink the time scale ${\tilde t}$ when shrinking the $\zeta$-plane,
but
for simplicity we denote the new time again by ${\tilde t}$.)

\subsubsection{Calculation of the winding angle distribution}

The imaginary (or vertical) coordinate $x$ in the rescaled $\zeta$-plane is
related to the
winding angle by $x=2\theta(t)/\ln t$. In order to obtain the winding angle
distribution, we have to determine the vertical position of a random walk
starting at the origin, at the moment when it reaches for the first time the
wall at distance one from the vertical  axis, without going beyond the
absorbing wall at distance $2|\ln a|/\ln t$ on the opposite side. Since we are
interested in a walk only up to the moment when it reaches the right-hand wall,
we can consider this wall also as absorbing. Since walks of length $t$ in the
original plane map on walks of different length $\tilde t$ in the new plane, we
need the probability that the walk is absorbed at this wall before time
${\tilde t}$.

We formulate this problem more generally and determine the probability
$P_{ \alpha, \beta}(y,{\tilde t})$ that a one-dimensional random walk
starting at $y$ between two absorbing points $\alpha$ and $\beta$  at time 0 is
absorbed at the point $ \beta$ before time ${\tilde t}$.  Since for
sufficiently small $\Delta {\tilde t}$, the walker is only a short distance
$\Delta y$ from its starting point, we have
\begin{eqnarray*}
 && P_{ \alpha, \beta}(y,{\tilde t}) = \int_0^\infty d(\Delta y) {1\over
\sqrt{2\pi\Delta
{\tilde t}}}
\exp\left[-{(\Delta y)^2\over 2\Delta {\tilde t}} \right]  \\
& &\times
\left[P_{ \alpha, \beta}(y + \Delta y,{\tilde t} - \Delta {\tilde t}) + P_{
\alpha, \beta}(y - \Delta y,
{\tilde t} - \Delta {\tilde t}) \right] \, .
\end{eqnarray*}
Expanding the above equation to the order of $\Delta \tilde t$ indicates that
$P_{ \alpha, \beta}(y,{\tilde t})$
satisfies a diffusion equation. The appropriate boundary conditions are
$P_{ \alpha, \beta}( \alpha,{\tilde t}) = 0$ and $P_{ \alpha, \beta}(
\beta,{\tilde t}) = 1$ with the initial value
$P_{ \alpha, \beta}(y,0) = 0$, resulting in\cite{kni81}
\begin{eqnarray*}
P_{ \alpha, \beta}(y,{\tilde t}) &=& {y -  \alpha \over  \beta -  \alpha} +
{2\over \pi}
\sum_{\nu = 1}^\infty {(-1)^{\nu + 1} \over \nu} \sin\left({\pi
\nu (y -  \alpha) \over  \beta -  \alpha}\right) \\
& & \quad \times
\exp\left[-{1\over 2} \left({\pi \nu \over  \beta -  \alpha}\right)^2 {\tilde
t}\right]\, .
\end{eqnarray*}
The probability  that the walk is absorbed at the right-hand boundary
during the time interval $\left[{\tilde t},{\tilde t} + d{\tilde t}\right]$, is
$ d{\tilde t}\, \partial_{\tilde t} P_{ \alpha, \beta}(y,{\tilde t}) $.

Note, however, that
$\int_0^\infty d{\tilde t}\,  \partial_{\tilde t} P_{ \alpha, \beta}(y,{\tilde
t}) =(y-\alpha)/(\beta-\alpha)$, i.e. equal
to the total fraction of particles absorbed at the right-hand boundary
(inversely proportional to the separations from the boundaries).
To calculate the winding angle distribution $p_A(x)$, we need the fraction of
these walks absorbed
between ${\tilde t}$ and ${\tilde t}+d{\tilde t}$, equal to
$((\beta-\alpha)/(y-\alpha)) \partial_{\tilde t} P$. Hence
(with $ \alpha = 2\ln a/ \ln t$, $ \beta = 1$ and $y = 0$)
\begin{eqnarray}
p_{A}(x) & = & \int_0^\infty d{\tilde t}\, {1 -  \alpha\over - \alpha}{\partial
P_{ \alpha,1}(0,{\tilde t}) \over \partial {\tilde t}}{\exp\left(-x^2/2{\tilde
t}\right) \over
\sqrt{2\pi{\tilde t}}}   \nonumber \\
&=& \int_0^\infty d{\tilde t} \, \sum_{\nu = 1}^\infty {(-1)^{\nu + 1} \over
\sqrt{2\pi{\tilde t}}}  {\pi \nu \over  \alpha(1 -  \alpha)} \sin\left( {\pi
\nu
\alpha\over 1 -  \alpha}\right)\nonumber \\
& & \quad \times \exp\left[-{1 \over 2} \left( {\pi \nu \over 1 -
 \alpha}\right)^2
 {\tilde t} - {x^2 \over 2{\tilde t}}\right] \nonumber \\
&=& \sum_{\nu = 1}^\infty {(-1)^{\nu + 1} \over  \alpha} \sin\left(
{\pi \nu  \alpha\over 1 -  \alpha}\right) \exp\left[-{\pi\nu |x| \over (1 -
 \alpha)}
\right] \, .\nonumber
\end{eqnarray}
The last step is achieved by first performing a Fourier transform with
respect to $x$, followed by integrating over ${\tilde t}$, and finally
inverting
the Fourier transform. (Alternatively, the ${\tilde t}$ integration can be
performed
by the saddle point method.) In the limit of large $t$, the variable $\alpha$
is very
small, and we can replace the sine--function by its argument.

Taking the sum
over $\nu$, we find
\begin{equation}
p_{A}(x) = {\pi \over  (1 -  \alpha)} {\exp\left[\pi x/(1 -  \alpha)\right]
\over \left\{\exp\left[\pi x/(1 -  \alpha)\right] + 1\right\}^2}\, .
\label{eqa}
\end{equation}
Changing the variable from $x$ to
\begin{displaymath}
\tilde x = {x\over (1 -  \alpha)} = {2\theta\over \ln\left(t/\alpha^2\right)}\,
,
\end{displaymath}
and noting that $p_{A}(x) dx = p_A(\tilde x) d\tilde x$, leads from
Eq.~(\ref{eqa})  to
\begin{equation}
p_A\left(\tilde x = {2\theta\over \ln\left(t/a^2\right)}\right)
= {\pi \over 4 \cosh^2(\pi\tilde x/2)}\, .
\label{abs}
\end{equation}

The above distribution,
which is exact in the limit $t \to \infty$, has an exponential decay
for large $\tilde x$, as first derived in Ref.~\cite{rud87}. The complete form
of Eq.~(\ref{abs}) was first given in Ref.~\cite{sal94}, however, without
derivation. The analogy to
random walkers in the plane $\zeta$, confined by the two walls,
provides simple physical justifications for the behavior of the
winding angle. In the presence of both walls, the diffusing
particle is confined to a strip, and loses any memory of  its starting
position at long times. The probability that a particle that has already
traveled a distance $\theta$ in the vertical direction proceeds a
further distance $d\theta$ without hitting either wall is thus
independent of $\theta$, leading to the exponential decay.

\subsubsection{Comparison to the winding angle distribution around a point
center}

The method described in this section was used earlier to derive the winding
angle distribution for Brownian motion  around a point center \cite{dur84}.
The resulting probability distribution for the winding angle in this case is
\cite{spi58}
\begin{equation}
\lim_{t\to\infty}p\left(x={2\theta \over\ln t}\right) = {1\over \pi}\,
{1 \over 1 + x^2}, \label{Cauchy}
\end{equation}
leading to an infinite mean winding angle.
Since there is  no confining wall on the left-hand side,
the particles may diffuse arbitrarily far in that direction, making it less
probable to hit the wall on the right hand side. In the original $z$-plane, the
walker takes no time at all to make an infinitely small winding around the
point center. This is clearly an unphysical feature, since real winding centers
are finite and since real random walks (or polymers) need a finite time (a
finite length segment) to make a winding. We therefore do not consider this
situation any further.

{\it Exercise: Derive Equation (\ref{Cauchy}), repeating the calculation of
this section, but with no absorbing wall (see Ref.~\cite{dro96}). }

\subsection{Winding of directed polymers around neutral winding centers}
\label{neutral}

Instead of having first a rod and then inserting the DP into the system, we can
also first have a free configuration of a DP, and then insert a rod into it. If
the polymer cannot relax to its thermal equilibrium distribution after
insertion of the rod, e.g. because its end are fixed or because its
configuration is frozen, the resulting winding angle distribution will be
different from that in the previous subsection. No configuration of the polymer
is forbidden, but those configurations that interfere with the rod become {\it
deformed}. The degree of deformation may depend on the diameter of the rod, but
the winding angle does not. Alternatively, we could consider a winding center
that has no interaction at all with the polymer, e.g. a light beam, or some
structural defect in the solvent that is not felt by the polymer. In this case
we would find the same winding angle distribution as in the case of a rod that
deforms the polymer.  In the language of a random walk, this situation
corresponds to having a disc that reflects all walks that hit it. The walks
that would go through the disc thus become deformed, but are not removed from
the statistical ensemble.

 We can obtain  the winding angle distribution by repeating the calculations
of the previous subsection, but replacing the absorbing boundary
condition $P_{ \alpha, \beta}( \alpha,{\tilde t}) = 0$ with the reflecting
condition
$\partial P_{\alpha, \beta}( y,{\tilde t})/\partial y\mid_{y=\alpha} = 0$,
leading to
\begin{eqnarray}
P_{\alpha,\beta}(y,\tilde t) &=& 1 - {2\over \pi} \sum_{\nu = 0}^\infty {1\over
\nu + 1/2} \sin\left(\pi ( \nu + 1/2) (\beta-y)\over \beta-\alpha \right)
\nonumber\\
&&\qquad \times \exp\left[-{1\over 2} \left({\pi (\nu + 1/2) \over  \beta -
 \alpha}
\right)^2 \tilde t\right]. \label{reflP}
\end{eqnarray}
There is thus no
current leaving the system at point $\alpha$, and walkers which hit the
winding center are reflected. We then find the winding angle distribution
\begin{equation}
p_{R}(\tilde x) =  {1\over 2\cosh(\pi\tilde x /2)}\,, \label{refl}
\end{equation}
where again $\tilde x = 2\theta/ \ln(t/a^2)$, and the limit
$t \to \infty$ has been taken.
For large $\tilde x$, where the walk has lost the memory of its initial
distance from both walls, this probability decays exponentially as
$\exp\left[-\pi\tilde{x}/2\right]$, i.e. exactly half as fast as for absorbing
boundary conditions. A random walk confined between an
absorbing and a reflecting wall that have a distance one can be mapped
to a random walk confined between two absorbing walls at distance two.
After rescaling the wall distance and the $\tilde x$-coordinate by two,
this explains the factor 1/2 between the decay constants in the tails
of the distributions in Eqs.~(\ref{abs}) and (\ref{refl}).

{\it Exercise: Derive Equation (\ref{refl}) (see Ref.~\cite{dro96}). }

\subsection{Universality of the winding angle distribution}
\label{univers}

The winding angle distribution in Eq.~(\ref{refl}), which we derived in the
previous subsection for Brownian motion around a reflecting disc, was obtained
previously by several authors in different contexts: B\'elisle \cite{bel89}
calculated the winding angle distribution for a random walk on a
two-dimensional lattice around a point that is different from any lattice site,
and for a random walk with steps of finite size taken in arbitrary directions,
around a point in the two-dimensional plane, obtaining in both cases
Eq.~(\ref{refl}).
The same result was obtained by Pitman and Yor ~\cite{pit86} for the
distribution of ``big windings'' of Brownian motion around two point-like
winding centers. Comtet, Desbois, and Monthus \cite{com93} divided the
two-dimensional plane into three concentric sections and determined the
contribution of each section to the windig angle for Brownian motion around a
point, finding Eq.~(\ref{refl}) for the contribution of the outer section.
This universality seems surprising, since one might expect that the main
increase in winding angle occurs when the walk is close to the winding center,
where details like the lattice symmetry, and shape and size of the winding
center determine how much time it takes to make one winding. However, a careful
look at Fig.~\ref{trafo} reveals that this is not the case: The main increase
in winding angle does not occur when the walker is within a small distance from
the left-hand wall. Since all distances have been scaled by $1/\ln(t)$, a small
distance from the left-hand wall corresponds to a large distance (of the order
of $\ln t$) from the winding center. Therefore, almost all windings are made
far from the winding center, where microscopic details do not matter. The
properties that do affect the winding angle distribution are conservation laws
(absorbing or reflecting boundary conditions), symmetries (with respect to the
sign of the angle - see the following subsection), singularities (as for the
winding of Brownian motion around a point center), and interactions
(self-avoidance) or randomness (see section \ref{nonideal}).

To further test this unversality hypothesis, we determined numerically the
winding angle distribution for a random walk on a lattice with reflecting and
absorbing boundary conditions. Reflecting boundary
conditions are realized by choosing a winding center different from
the vertices of the lattice, and thus never crossed by the walker (this is
exactly the situation treated analytically in Ref.~\cite{bel89}).
On the other hand, to model absorbing boundary conditions, the winding
center  is chosen as one of the lattice sites (say the origin),
but no walk is allowed to go through this point.

The winding angle distributions are most readily obtained using a transfer
matrix method which calculates the number of all walks with given winding
angle and given endpoint after $t$ steps, from the same information after
$t-1$ steps. The winding center is at $(0.5,0.5)$ for reflecting boundary
conditions, and at the origin for absorbing boundary conditions.
The walker starts at (1,0), and the winding angle is increased or
decreased by $2\pi$ every time it crosses the positive branch
 of the $x_1$--axis.
Due to limitations in computer memory,
we applied a cutoff in system size and winding angle for times $t > 120$,
making sure that the results are not affected by this approximation. The
largest
times used, $t=9728$, required approximately 3 days to run on a Silicon
Graphics Indy Workstation.

Figures \ref{squarerefl} and
\ref{squareabs} show the results for the two cases. The asymptotic
exponential tails predicted by theory can clearly be seen; deviations from
the theoretical curve for smaller values of the scaling variable $x =
2\theta/\ln(2t)$
are due to the slow convergence to the asymptotic limit.  Since the scaling
variable depends logarithmically on time, the asymptotic limit is reached only
for large $\ln t$. Note that the only free parameter in fitting to the
analytical form
is the characteristic time scale appearing inside the logarithm. With $t$
measured in units of single steps on the lattice, we found that a factor 2 in
the
scaling variable provides the best fit. In the limit $t \to \infty$, different
scales of
$t$ give of course the same asymptotic winding angle distribution.

{\it Exercise: Perform the numerical calculations mentioned in this section.
Study also the case of absorbing and reflecting winding centers that comprise
several lattice points. How does the size of the winding center affect the
convergence towards the asymptotic winding angle distribution?}

We also studied the winding of a DP proceeding along the diagonal
of a cubic lattice in three dimensions (see Fig.~\ref{3dlattice}). The polymer
starts at $(1,0,0)$, and at each
step increases one of its three coordinates by 1. We determined the
winding angle distribution around the diagonal $(1,1,1)$-- direction,
excluding from the walk all points that are on this diagonal (a repulsive
columnar defect, corresponding to the case of an absorbing winding center).
The excluded points lie on the origin when the polymer is projected in a plane
perpendicular to the diagonal. In this plane, the polymer proceeds along the
bonds of a triangular lattice, alternating between the three different
sublattices. A cutoff of $243$ in system size was imposed for
the transfer matrix calculations. The winding angle distribution
$p(x = 2\theta/\ln t)$ is shown in Fig.~\ref{fig1} for different times. As for
the square lattice, an exponential tail with decay constant of $\pi$ can
be seen. Our numerical results, as well as the analytical considerations,
thus indicate clearly that the winding angle distributions for reflecting and
absorbing boundary conditions are universal and do not depend on
microscopic details.

Due to the special properties of directed paths along the diagonal of the
cube, the case of reflecting boundary conditions leads to an asymmetry between
windings in positive and negative directions. This is because it takes only
three steps to make the smallest possible winding in one direction, but six
steps in the opposite direction. This situation is discussed in detail in
the following subsection.

\subsection{Winding centers with chiral asymmetry}
\label{chiral}

So far, we only considered situations that are symmetric with respect to the
angles
$\pm\theta$. For directed paths on certain lattices, however, this symmetry is
broken. A directed walk that proceeds at each
step along $+x_1$, $+x_2$, or $+x_3$ direction on a cubic lattice can be mapped
on a
random walker on a two-dimensional triangular lattice as indicated in figure
\ref{gitter}(a).
Each bond can be crossed only in one direction,  and the winding center for
reflecting boundary conditions must be different from the vertices of the
lattice. It is apparent from this figure that the random walker can go around
the center in 3 steps in one angular direction, but in no less than 6 steps
in the other direction.

An alternative description is obtained by examining the position of the walker
after every three time steps. The resulting coarse-grained random walk takes
place on a regular triangular lattice, but now the walker has a finite
probability of
$3\times 2/3^3=2/9$ of staying at the same site. If this site is one of the
three
points next to the winding center, the winding angle is increased by $2\pi$
in one of the six possible configurations that return to the site after three
steps.
In other words, the walker has a finite probability of having its winding angle
increased in the proximity of the center. The amount of this biased increase in
angle depends on the structure of the lattice and will be different
for other directed lattices. An equivalent physical situation occurs
for Brownian motion around a rotating winding center, e.g. a rotating
reflecting
disc which does not set the surrounding gas or liquid into motion (see
Fig.~\ref{rotate}).

Since this angular symmetry breaking is already present in the above simple
example of a directed walk, it is quite likely to occur in  more realistic
physical
systems, such as with screw dislocations in the underlying medium. The winding
angle distribution for two {\it chiral} polymers \cite{deGen82} should also
show an angular asymmetry.
We thus use the term chiral winding center to indicate that
each time the polymer comes close to the center, it finds it easier to wind
around
in one direction as opposed to the other. (Of course, to respect the reflecting
boundary conditions, there must be no additional interaction with the winding
center. The case of additional attractive or repulsive interaction with the
winding center will be discussed later.)

After mapping to the rescaled $\zeta$-plane introduced in Sec.~\ref{free},
the above situations can be modeled by a downward moving, reflecting wall
on the vertical axis. Each time a random walker hits this wall, its vertical
position $x = 2\theta/\ln t$ is changed by a small amount $2\Delta\theta/\ln
t$.
Let us now determine the net shift $\Delta x$ in $x$ due to the motion of the
wall for a
walker that survives for a time ${\tilde t}$ in the rescaled $\zeta$-plane,
before it is
absorbed at the right-hand wall (recall that $t$ is the time in the original
system,
while ${\tilde t}$ refers to the time in the rescaled $\zeta-$plane, after the
conformal
mapping).

To obtain the full solution, it is necessary to solve the two-dimensional
diffusion
equation with moving boundary conditions. Since we are mainly interested in
the exponential tails of the winding angle distribution, we restrict our
analysis
to the limit of large times ${\tilde t}$, and determine the shift in $x$ due to
encounters with the reflecting moving wall in this limit. A Brownian walker
that has survived for a sufficiently long time ${\tilde t}$ forgets its initial
horizontal
position. The mean number of encounters with the reflecting wall, and
consequently the shift $\Delta x$  in $x$ due to the motion of the wall, is
then
expected to be simply proportional to the considered time interval. Applying
the central limit theorem in the limit ${\tilde t} \to \infty$, the
probability distribution of $\Delta x$  is given by
\begin{equation}\label{pDelta}
p_\Delta(\Delta x)={1\over\sqrt{2\pi\beta^2{\tilde t}}}
\exp\left[-{(\Delta x -\alpha {\tilde t})^2\over 2\beta^2 {\tilde t} }\right].
\end{equation}
The parameters $\alpha$ and $\beta$ are related to the velocity of the wall
(chirality of the defect) by $\alpha\propto\beta\propto v$. Presumably
Eq.~(\ref{pDelta}) can be obtained directly from properties of random walks,
providing the exact coefficients for these proportionalities.

The tail of the winding angle distribution is then given by
\begin{eqnarray}
p^c_{R}(x) & \propto & \int_0^\infty d{\tilde t} \int_{-\infty}^{\infty}
d(\Delta x)  {\partial P_{ \alpha,1}(0,{\tilde t}) \over \partial {\tilde t}}
 {1 \over \sqrt{2\pi{\tilde t}}}
\exp\left[-{(x-\Delta x)^2\over 2 {\tilde t}}\right] \nonumber \\
 & \propto &  \int_0^\infty d{\tilde t}
\exp\left[-(\pi^2/4) \tilde t / 2\right]
{1\over \sqrt{2\pi{\tilde t}(1+\beta^2)}}
\exp\left[-{(x-\alpha {\tilde t})^2\over 2{\tilde t}(1+\beta^2)}\right]
\nonumber \\
&= &\exp\left[{\alpha x\over 1+\beta^2}
-{|x|\over 1+\beta^2}\sqrt{\alpha^2+{\pi^2\over 4}
\left(1+\beta^2\right)}\right], \label{chir}
\end{eqnarray}
valid for large $|x|$. The second factor on the right-hand side of the first
line of Eq.~(\ref{chir})
is  the probability that the random walker is absorbed at the right-hand wall
at time $\tilde t$ (see Eq.~(\ref{reflP}), and is for large $\tilde t$
dominated by the slowest mode ($\nu = 0$). The second factor is the probability
that the random walk would have the vertical coordinate $x$ after time $\tilde
t$, if there were no motion of the wall.
The effect of the moving wall on the winding angle distribution is thus a
systematic shift in the slopes of the exponential tails. For small values of
chirality the slopes on the two sides are changed to $\pi/2\pm \alpha$.
Due to this explicit velocity dependence, these asymmetric distributions are
clearly non-universal. At large chiralities, the slopes vanish as
$\alpha/\beta^2$ resulting in quite wide distributions. Apparently strong
chirality of a defect increases the probability of entanglements.
Fig.~\ref{chiral1} shows our simulation results for the winding angle
distribution for a walk on the above mentioned directed triangular lattice.
The asymmetry due to the shift is clearly visible, and the winding angle
distribution is wider than for a stationary wall. This case thus exemplifies
the strong chirality limit discussed in the previous paragraph.
We also simulated a square lattice with directed bonds as indicated in
Fig.~\ref{gitter}(b). The corresponding winding angle distribution is shown in
Fig.~\ref{chiral2}. The distribution is again asymmetric, but not as wide
as in the previous case, and more similar to that expected in the weak
chirality regime.

So far, we have assumed that the winding angle changes by the same amount each
time the walker returns to the winding center. It is more realistic to assume
that the change in winding angle has a certain probability distribution. A
possible example is provided by polymers with randomly changing chirality
\cite{sel96}.
In the limit $t\to \infty$, the total change in winding angle due to the
chirality is on an average
$$\langle \Delta \theta \rangle = \int_{-\infty}^\infty d(\Delta \theta)\,
p(\Delta \theta) \Delta \theta, $$
where $p(\Delta \theta)$  is the probability distribution for $\Delta \theta$.
The variance for the total change in the scaling variable $x$ is
$$
(\delta\Delta x)^2 \simeq \ln t \left(2\delta\Delta
\theta\over \ln t \right)^2
$$
and vanishes in the limit $t \to \infty$. The effect of random chirality of the
winding angle distribution is identical to that of uniform chirality, and is
zero when segments of positive and negative chirality occur equally often.

Finally, we want to emphasize that the results of this section are based on the
assumption that there is no interaction between the polymer and the winding
center besides the chirality. When there is an additional repulsive
interaction, we have to choose absorbing boundary conditions, in which case the
random walk never hits the moving wall, and the motion of the wall (the
chirality) has no effect at all on the winding angle distribution. On the other
hand, when the polymer is bound to the winding center due to an attractive
interaction, the number of returns to the winding center,  and consequently the
systematic shift in the winding angle due to chirality are proportional to the
length $t$ (see section \ref{confine}). In the presence of both an attractive
interaction and a hard-core repulsion (probably the most realistic case
\cite{HLV}), the polymer performs a phase transition from a bound to a free
state depending on the temperature, and we expect the results of this
subsection to apply near the transition temperature.

\section{The winding angle distribution of confined polymers}
\label{confine}

Up to now, we considered only cases where the polymer could wander infinitely
far away from the winding center. However, there are many physical situations
where polymers indeed are confined to some region around the winding center.
When the diameter of the container is small compared to the length of a DP, or
when the polymer density is so large that for each of them just a small
cylinder is available, the winding angle distribution will be fundamentally
different from the previous section. Random walk segements of time $\Delta t$
that are small compared to the total length of the walk, but large compared to
the time it takes to make a winding, have identical winding angle
distributions. Applying the central limit theorem, we can therefore predict
that the total winding angle distribution of a confined polymer has the form
\begin{equation}
p_{con}(\theta) \propto \exp\left[-a\theta^2/2t\right]\, ,
\end{equation}
where $1/a$ is the variance in the winding angle per unit time.

When the winding center is chirally asymmetric, there is an additional shift in
the mean winding angle. In the following three subsections, we will determine
the winding angle distribution for polymers confined between two cylinders,
polymers bound to an attractive winding center, and bound polymers winding
around chiral centers.

\subsection{Polymer confined between two cylinders}
\label{confine1}

We start with the simple model of a polymer confined between two concentric
cylinders (see Fig.~\ref{cylinder}). The inner cylinder is the winding center,
the outer one is the wall of the container, or represents the repulsion of the
neighboring polymers. This situation is equivalent to a random walk confined
between two concentric rings of radii $R_1$ and $R_2$.
 After a long time, the probability $p(r)$ to find the walker at a given radius
$r$ is independent of time, and of the angle. For reflecting boundary
conditions at the outer and the inner ring, the walker is then with equal
probablity at any site between the two rings, leading to
$$p(r) = {1 \over r  \ln(R_2/R_1)}\,.$$
The variance of the winding angle per unit time is $1/r^2$ when the walker is
at radius $r$. Since the variances of different segments of the random walk are
independent, we can add them up,  leading in the limit of large $t$ to
$${1\over a} = \int_{R_1}^{R_2} {1\over r^2} p(r) dr = \int_{R_1}^{R_2}  {dr
\over r^3  \ln(R_2/R_1)}
= {t\over 2\ln(R_2/R_1)}\left({1\over R_1^2} - {1 \over R_2^2}\right)\,.$$

It is more physically relevant to use absorbing boundary conditions with  $p(r)
= 0 \hbox{ for } r = R_1,R_2$. We solve the diffusion equation
$$ {\partial P(r,\phi,t) \over \partial t} = {1\over 2}\left( {\partial^2 P
\over \partial r^2} + {1\over r} {\partial P \over \partial r} + {1\over r^2}
{\partial P \over \phi^2}\right)$$
with the ansatz
$$P(r,\phi,t) = \sum_{n=0}^\infty p(r) \cos(n\phi) \exp\left[-\lambda_n
t\right].$$
For long times, the angular dependence vanishes, and the mode with the slowest
decay (the smallest eigenvalue $\lambda_0$) dominates. (Note that $\phi$ is not
the winding angle, but the azimutal angle, which takes only values between 0
and $2\pi$.) Since we normalize the winding angle distribution with respect to
the  walkers that do not get absorbed, the factor $\exp[-\lambda_0 t]$ drops
out, and the probability to find the walker after time $t$ at radius $r$ is
given by the solution of
$${\partial^2 p(r) \over \partial r^2} + {1\over r} {\partial p(r) \over
\partial r} + 2\lambda_0 p(r) = 0,$$
with the boundary conditions given above, and with the normalization condition
$\int_{R_1}^{R_2} p(r) dr = 1$.
The general solution of this (Bessel) differential equation can be written in
form of an integral
$$ p(r) = \int_0^\pi \left[C_1 \cos\left(\sqrt{2\lambda_0}r\sin \zeta\right) +
C_2 \cos\left(\sqrt{2\lambda_0}r\cos \zeta\right) \ln\left(\sqrt{2\lambda_0}r
\sin^2 \zeta\right) \right] d\zeta. $$
The values of $\lambda_0$, $C_1$, and $C_2$ are obtained by matching two
consecutive zeros of this function to $r=R_1$ and $r=R_2$, and by normalizing
properly. In general, the solution cannot be written down in a closed form and
has to be found numerically. In the case $R_1 \ll R_2$, the values of
$\lambda_0$ and $C_1/C_2$ can be found analytically, since $C_1 \gg C_2$ in
this limit, and the first two zeros of $p(r)$ are given by the conditions
$\ln(\sqrt{2\lambda_0}R_1) = C_1/C_2$ and $\sqrt{2\lambda_0} R_1 \simeq 2.4 $
(the first zero of the Bessel function $J_0$).
In the limit $R_2-R_1 \ll 1$, we find $p(r) = C \sin\left(\pi {r-R_1 \over
R_2-R_1}\right)$, where $C$ is the normalization constant.

{\it Exercise: Set $R_1 = 1$ and determine numerically the variance
$${1\over a} =  \int_{R_1}^{R_2}  {p(r)dr \over r^2 }$$
as a function of $R_2$. }

\subsection{Polymer bound to an attractive winding center}

A polymer can also be confined by an attractive winding center (see
Fig.~\ref{attractive}):
A DP subject to an attractive potential of radius $b_0$
and binding energy $U_0$ per unit length, is bound to that winding center. For
temperatures above the crossover value of $T^* \propto b_0 \sqrt{U_0}$,
the polymer is only weakly bound and wanders horizontally over a large
localization length, $l_{\perp}(T) \simeq b_0 \exp[(T/T^*)^2]$\cite{nel93}.
The mean vertical distance $l_z$ between consecutive intersections of the
polymer with the defect is consequently proportional to $l_{\perp}^2$. Over
this
distance, the polymer can be approximated by a directed walk which returns to
its
starting point (the winding center) after a time $l_z$.

Using the result in Eq.~(\ref{abs}), we can derive the winding angle
distribution $p_A^o(\tilde x)$ for such confined random walks. Each walk that
returns to its starting point after time $t$ (in the $z$-plane) is composed of
two
walks of length $t/2$ going from the starting point to $z(t/2)$. As we have
seen
in subsection \ref{absorbing}, almost all walks of length $t/2$, when mapped on
the plane $2\zeta/\ln(t/2)$, have their endpoint on a vertical wall at distance
one from the origin. The winding angle distribution for these walks is given
by Eq.~(\ref{abs}), with $t$ replaced by $t/2$. The probability that a walk
that returns to its starting point has a winding angle $\theta$ is therefore
obtained by adding the probabilities of all combinations of two walks of length
$t/2$ whose winding angles add up to $\theta$, i.e.
\begin{eqnarray}
p_A^o(\tilde x) &=& \int_{-\infty}^\infty dy {\pi\over 4 \cosh^2(\pi y)}
{\pi\over 4 \cosh^2\left(\pi(\tilde x -  y )\right)}\nonumber\\
&=& {\pi\over2 \sinh^2(\pi\tilde x/2)}\left({\pi \tilde{x} \over 2}
\coth({\pi \tilde{x} \over 2}) - 1\right) \,, \label{abso}
\end{eqnarray}
where $\tilde x = 2\theta / \ln(t/2R^2)$.

For large $\tilde{x}\approx x$, the above expression decays as
$x\exp\left[-\pi x\right]$.
A polymer of length $t$ is roughly broken up into $t/l_z$ segments between
contacts with the attractive columnar defect. We can assume that the winding
angle of each segment is independently taken from the probability distribution
in Eq.~(\ref{abso}) with $t\approx l_z$. Adding the winding angle distributions
of all segments leads to a Gaussian distribution centered around $\theta = 0$,
and with a variance proportional to $L\ln(l_z)/l_z$.

\subsection{Confined polymers winding around chiral centers}

When the winding center has a chiral asymmetry, the mean winding angle is
increased by some finite amount $\Delta \theta $ per unit time. The winding
angle distribution is consequently modified to
$$p_c(\theta) = \exp\left[- \tilde a(\theta - t\Delta \theta)^2/t\right],$$
with a mean winding angle proportional to the length of the polymer (see also
\cite{HLV}).
$1/\tilde a$ is larger than $1/a$, since the variances in the number of returns
to the winding center, and in $\Delta \theta$, both contribute to the variance
of the winding angle. For weak chirality, $1/\tilde a$ is close to $1/a$, and
the main effect of the chirality is just a shift of winding angle distribution.
For strong chirality, we expect $1/\tilde a \gg 1/a$, and the winding angle
distribution becomes very broad (similar to the situation discussed in  section
\ref{chiral}).

When the polymer is confined not by a container but by neighboring polymers, it
will not just wind around one of these neighbors. Kamien and Nelson
\cite{kam95} have shown that when chirality is strong, screw dislocations
proliferate throughout the polymer crystal.

\section{Winding angles in random media and for self-avoiding polymers}
\label{nonideal}

So far, we only considered winding topologies that can be mapped on ideal
random walks. However, when the medium in which the polymer is embedded is non
homogeneuos, the energies of different polymer configurations are different.
Examples are polymers in gels and porous media \cite{foo96}, or magnetic flux
lines in high-$T_c$ superconductors, pinned by oxygen impurities \cite{HTSC}.
We consider the case of {\it quenched} randomness, where one end of the polymer
is fixed \cite{dou91}.

The behavior of a DP in the presence
of short-range correlated randomness is modeled by a directed path on a lattice
with random
bond energies\cite{kardarrev}. In 3 or less dimensions, the polymer is always
pinned at
sufficiently long length scales. An important consequence of the pinning is
that the
path wanders away from the origin much more than a random walk, its transverse
fluctuations scaling as $t^\nu$, where $\nu\approx 0.62$ in three dimensions,
and $\nu=2/3$ in two dimensions\cite{AmarFamily,KimBrayMoore}.
The probability of such paths returning to the winding center are thus greatly
reduced, and the winding probability distribution is expected to change.

We examined numerically the windings of a directed path along the diagonal
of a cubic lattice (see Fig.~\ref{3dlattice}). To each bond of this lattice was
assigned
an energy randomly chosen between 0 and 1. Since the statistical properties
of the pinned path are the same at finite and zero temperatures, we determined
the winding angle of the path of minimal energy by a transfer matrix method.
For each realization of randomness, this method\cite{kardarrev} finds the
minimum energy of all paths terminating at different points, and with different
winding numbers. This information is then updated from one time step to the
next. From each realization we thus extract an optimal angle as a function of
$t$. The probability distribution is then constructed by examining 2700
different
realizations of randomness. To improve the statistics, we averaged over
positive
and negative winding angles.

The resulting distribution is shown in figure \ref{fig3a}, with a scaling
variable $x = \theta/2\sqrt{\ln t}$.
This scaling form is motivated by that of self-avoiding walks, which in two
dimensions follow a Gaussian distribution
\begin{equation}\label{pSAW}
p_{SA}\left(x = {\theta\over 2\sqrt{\ln t}}\right)=
{1\over \sqrt{\pi}}\exp\left(-x^2\right).
\end{equation}
The result of the data collapse in Fig.~\ref{fig3a} agrees well with
the Gaussian distribution
\begin{equation}
p_{\rm rand}\left( x= {\theta\over 2\sqrt{\ln t} }\right)=
\sqrt{1.5\over \pi}\exp\left(-1.5x^2\right) \, .
\label{gauss}
\end{equation}
Directed paths in random media and self-avoiding walks share a number of
features which make the similarity in their winding angle distributions
plausible.
Both walks meander away with an exponent larger than the random walk
value of 1/2. (The exponent of 3/4 for self-avoiding walks is larger than
$\nu\approx 0.62$ for polymers in 3 dimensions.) As a result, the probability
of returns to the origin is vanishingly small in the limit $t\to\infty$ for
both types of
paths, and the properties of the winding center are expected to be
irrelevant. (A simple scaling argument suggests that the number of returns
to the origin scale as $N(t)\propto1/t^{1-2\nu}$.)
The conformal mapping of section \ref{free} cannot be
applied in either case: The density and size of impurities in a random medium
become coordinate dependent under this mapping, as does the excluded
volume effect. The winding angle distribution for self-avoiding walks in
Eq.~(\ref{pSAW}) has been
calculated using a more sophisticated mapping\cite{dup88,sal94}.
As a similar exact solution is not currently available for polymers in random
media,
we resort to the scaling argument presented next.

Let us divide the self-avoiding walk, or the directed path, in segments going
from
$t/2$ to $t$, from $t/4$ to $t/2$, etc., down to some cutoff length of the
order of the
lattice spacing, resulting in a total number of segments of the order of $\ln
t$ (see Fig.~\ref{scalingarg}).
The statistical self-similarity of the walks suggests that a segment of length
$t/2^n$ can be mapped onto a segment of length $t/2^{n+1}$ after rescaling by
a factor of $1/2^\nu$. Under this rescaling, the winding angle is
(statistically
speaking) conserved, and consequently all segments have the same winding
angle distribution. Convoluting the winding angle distributions of all
segments,
and assuming that the correlations between segments do not invalidate the
applicability of the central limit theorem, leads to a Gaussian distribution
with
a width proportional to $\ln t$. This argument does not work for the random
walks
considered in section \ref{free}, since the finite radius of the
winding center is a relevant parameter. Different segments of the walk are
therefore not statistically equivalent, as they see a winding center of
different
radius after rescaling.

 In the pure system, we had to distinguish between
repulsive, neutral, and chiral winding centers. Since in the presence of point
impurities the polymer does not return to the winding center as often, these
differences are now irrelevant.
In fact, it can be shown that even an attractive force between the winding
center  and the polymer cannot bind the polymer to the winding center, as long
as it does not exceed some finite threshold \cite{Bal94}.
Perhaps not surprisingly, the main conclusion of this section is that the
pinning to point
randomness decreases entanglement.

\section{Discussion and Conclusions}
\label{concl}

Topological entanglements present strong challenges to our understanding
of the dynamics of polymers and flux lines. In this paper, we examined the
windings
of a single directed polymer around a columnar winding center, or the winding
of two DPs around each other. By focusing on even this simple
physical situation we were able to uncover a variety of interesting properties:
The probability distributions for the winding angles can be classified into
a number of universality classes characterized by the presence or absence
of underlying symmetries or boundary conditions.

For free DPs in a homogeneous medium, we find a  number of exponentially
decaying distributions: If there is no interaction at all between the polymer
and the winding center (corresponding to reflecting boundaries for random
walks) we obtain the distribution in Eq.~(\ref{refl}). Removing this
conservation
(absorbing boundaries or repulsive interaction between the polymer and the
winding center) leads to the distribution in
Eq.~(\ref{abs}) whose tails decay twice as fast.

A completely new set of distributions is obtained for chirally asymmetric
situations, where the
polymer is preferentially twisted in one direction at the winding center. These
distributions have asymmetric exponential tails, with decay constants that
depend
on the degree of chirality. Strong chirality appears to lead to quite broad
distributions. A remaining challenge is to find the complete form of this
probability distribution by solving the two dimensional diffusion equation
with moving boundary conditions.

When the polymer is confined to a finite volume around the winding center, the
winding angle distribution becomes Gaussian, with a width proportional to the
length of the polymer. In the presence of chiral asymmetry, the mean winding
angle is proportional to the length of the polymer, and we have to distinguish
again between the limits of weak and strong chirality.

For non-ideal walks, with a vanishing probability to return to the origin, the
properties of the winding center are expected to be irrelevant. Both
self-avoiding
walks in $d=2$ dimensions, and polymers pinned by point impurities in $d=3$,
have wandering exponents $\nu$ larger than 1/2 and fall in this category.
We present a scaling argument (supported by numerical data) that in this case
the probability distribution has a Gaussian form in the variable
$\theta/\sqrt{\ln t}$.
Not surprisingly, wandering away from the center reduces entanglement.
The characteristic width of the Gaussian form is presumably a universal
constant that has been calculated exactly for self-avoiding walks in $d=2$.
It would be intersting to see if this constant (only estimated numerically for
the impurity pinned polymers in $d=3$) can be related to other universal
properties
of the walk. Changing the correlations of impurities (and hence the exponent
$\nu$) may provide a way of exploring such dependence.

There are certainly other universality classes not explored in this paper.
For example, we did not consider the case of a long-range interaction between
the polymer and the winding center. Also, the mapping of a DP in a nematic
solvent on a random walk is correct only to first approximation
\cite{nel94,deGen82}. Due to long-range correlations within the nematic
solvent, the number of returns of the polymer to the winding center does not
increase with $\ln t$, but with $\ln(\ln t)$ for very large $t$.

The results of this paper also provide conjectures for the winding of non
directed polymers around a rod. If the self-interaction of the polymer can be
neglected for some reason, the results of section \ref{free} can be applied,
the parameter $\tau$ now being the internal coordinate of the polymer. To first
approximation, this might be correct for a polymer close to the $\theta$-point,
but three-point interactions which ultimately swell the $\theta$-polymer will
eventually invalidate the result \cite{doi86}.

When the polymer swells to give $\nu > 1/2$, as is the case for a self-avoiding
random walk in three dimensions, the winding center is no longer important. The
projection into the plane perpendicular to a rod is a walk that wanders away
from the winding center faster than an ideal random walk. We can therefore
apply the results of section \ref{nonideal} and conclude that the winding angle
distribution is a Gaussian, with a variance proportional to $\ln t$. When the
polymer is in the collapsed state, its winding angle distribution is again
different, and has still to be found.  Since collapsed polymers are relatively
compact, they can be approximated by Hamiltonian walks that visit each site
within a volume of the size of the polymer exactly once \cite{orl85,pan94}. We
thus conclude this review with the open problem of determining the probability
distribution for windings in the collapsed state by  examining the behavior of
Hamiltonian walks.

\acknowledgements
We thank M.E. Fisher, A. Grosberg, Y. Kantor, P. LeDoussal, and S. Redner 
for helpful discussions.  BD is supported  by the  Deutsche 
Forschungsgemeinschaft (DFG) under Contract No.~Dr 300/1-1. MK  
acknowledges support from NSF grant number DMR-93-03667.

\begin{figure}
\caption{a) Two directed polymers winding around each other. b) A directed
polymer winding around a rod.
}
\label{picture}
\end{figure}

\begin{figure}
\caption{Conformal mapping of the random walk with absorbing boundary
conditions, and subsequent rescaling.
}
\label{trafo}
\end{figure}

\begin{figure}
\caption{Winding angle distribution for random walks on a square lattice
with {\it reflecting} boundary conditions for $t = 38$ (dot-dashed),
152 (long dashed),  608 (dashed), 2432 (dotted), and 9728 (solid).
The horizontal axis is $x=2\theta/\ln(2t)$. The thick solid line is the
analytical  result of Eq.~(\protect\ref{refl}).
}
\label{squarerefl}
\end{figure}

\begin{figure}
\caption{Winding angle distribution for random walks on a square lattice
with {\it absorbing} boundary conditions. The symbols and the variable
$x$ are the same as in the previous figure. The thick solid line is the
analytical result of Eq.~(\protect\ref{abs}).
}
\label{squareabs}
\end{figure}

\begin{figure}
\caption{a) Construction of a polymer directed along the (1,1,1)-diagonal of a
cubic lattice. b) projection into the plane perpendicular to the
(1,1,1)-diagonal. The three sublattices are indicated by different shades of
gray.
}
\label{3dlattice}
\end{figure}

\begin{figure}
\caption{Winding angle distribution around the preferred direction for a
flux line (directed path) in 3 dimensions for $t = 243$ (dot-dashed), 729
(long dashed),  2187 (dashed), 6561 (dotted),  and 19684 (solid).
The horizontal axis is $x=2\theta/\ln(2t)$. The thick solid line is the
analytical result of Eq.~(\protect\ref{abs}).
}
\label{fig1}
\end{figure}

\begin{figure}
\caption{Triangular and square lattices with directed bonds.
The winding centers are indicated by a circle o.
}
\label{gitter}
\end{figure}

\begin{figure}
\caption{Brownian motion around a rotating winding center.
}
\label{rotate}
\end{figure}

\begin{figure}
\caption{Winding angle distribution around the preferred direction for a
random walk on a directed triangular lattice for $t = 243$ (dot-dashed), 729
(long dashed),  2187 (dashed), 6561 (dotted), and 19684 (solid). The
scaling variable is $x=2\theta/\ln(2t)$. The thick solid line is the
distribution  given in Eq.~(\protect\ref{refl}).
}
\label{chiral1}
\end{figure}

\begin{figure}
\caption{Winding angle distribution for a random walk on a directed square
lattice for  $t = 38$ (dot-dashed), 152 (long dashed),  608 (dashed), 2432
(dotted), and 9728 (solid). The thick solid line is the
distribution  given in Eq.~(\protect\ref{refl}).
}
\label{chiral2}
\end{figure}

\begin{figure}
\caption{A directed polymer confined between two cylinders.
}
\label{cylinder}
\end{figure}

\begin{figure}
\caption{A directed polymer bound to an attractive rod.}
\label{attractive}
\end{figure}

\begin{figure}
\caption{Winding angle distribution for a directed path in a random
3-dimensional system for  $t = 120$ (dotted), 240 (dashed),
480 (long dashed), 960 (dot-dashed),  and 1920 (solid).
The thick solid line is the  Gaussian distribution in
Eq.~(\protect\ref{gauss}).
}
\label{fig3a}
\end{figure}

\begin{figure}
\caption{Division of the self-avoiding walk into self-similar segments.
}
\label{scalingarg}
\end{figure}

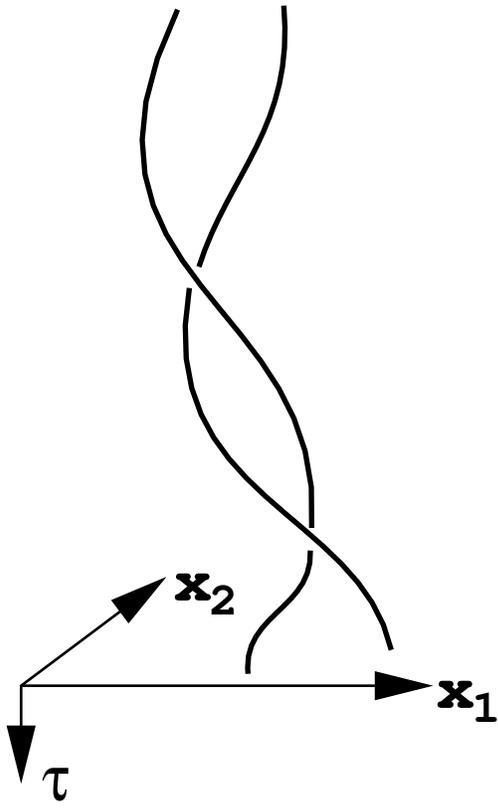
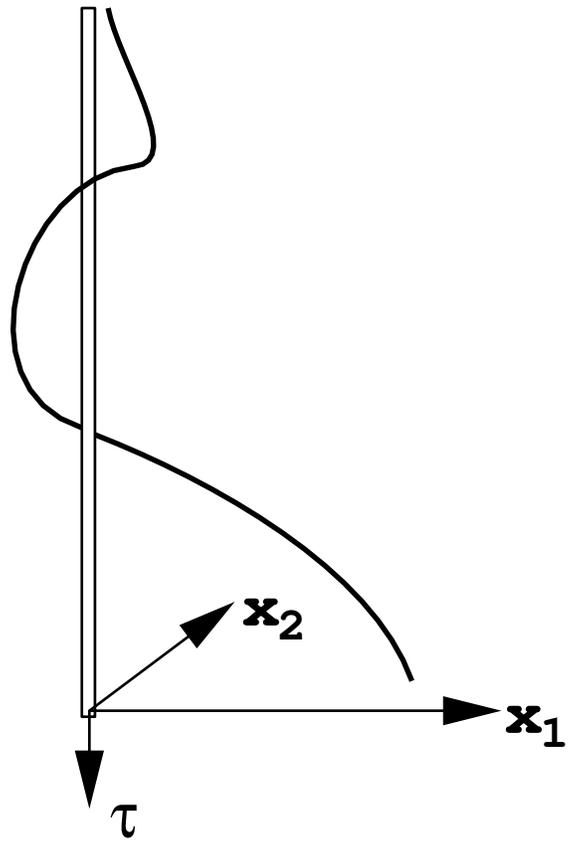
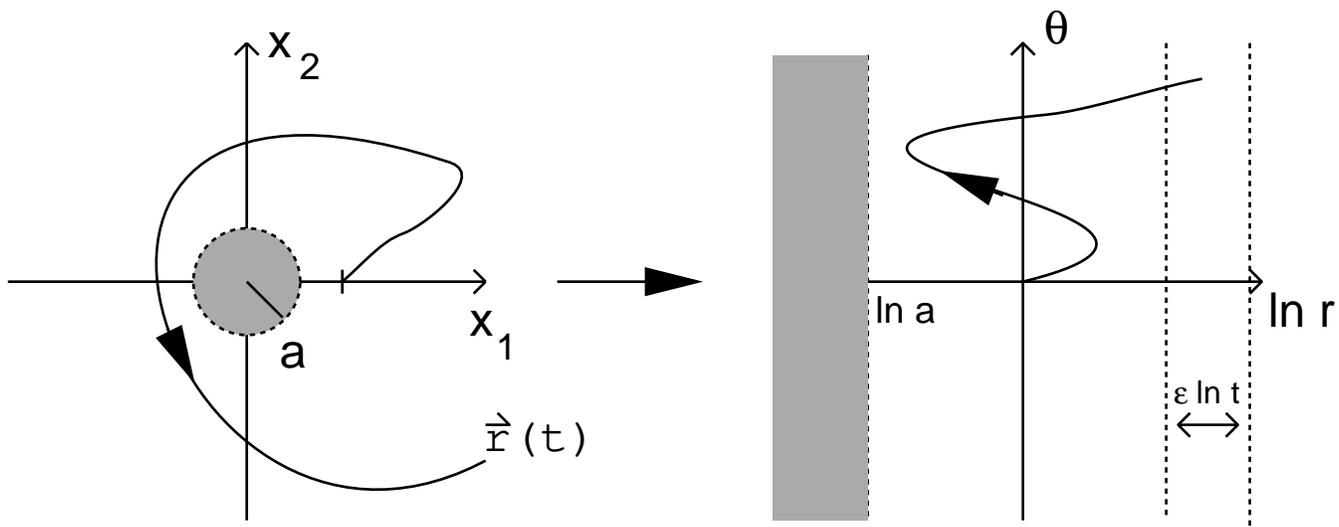
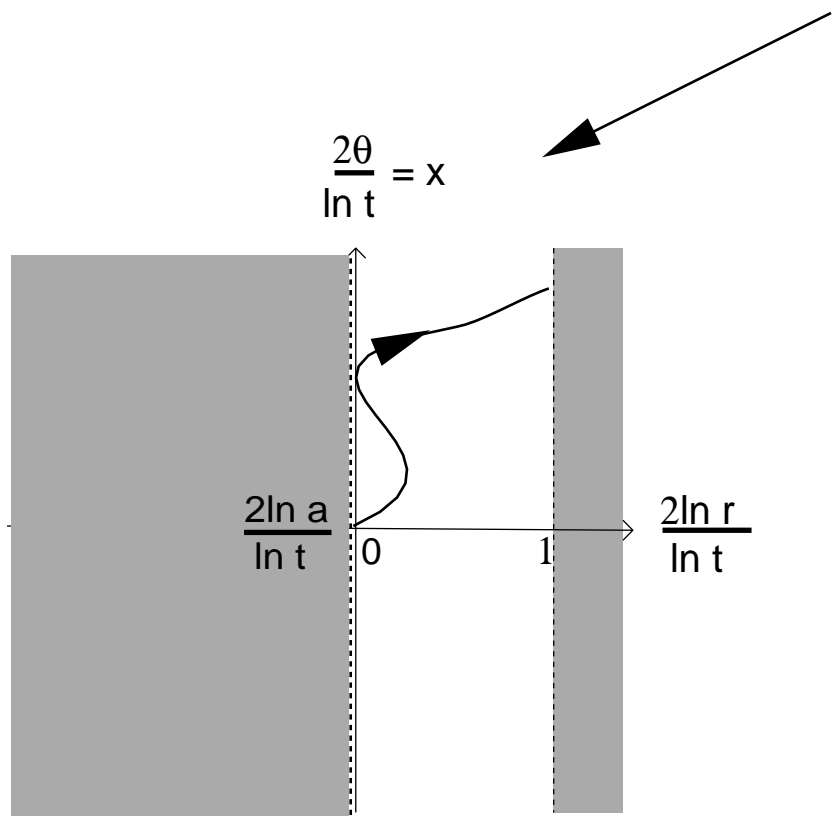
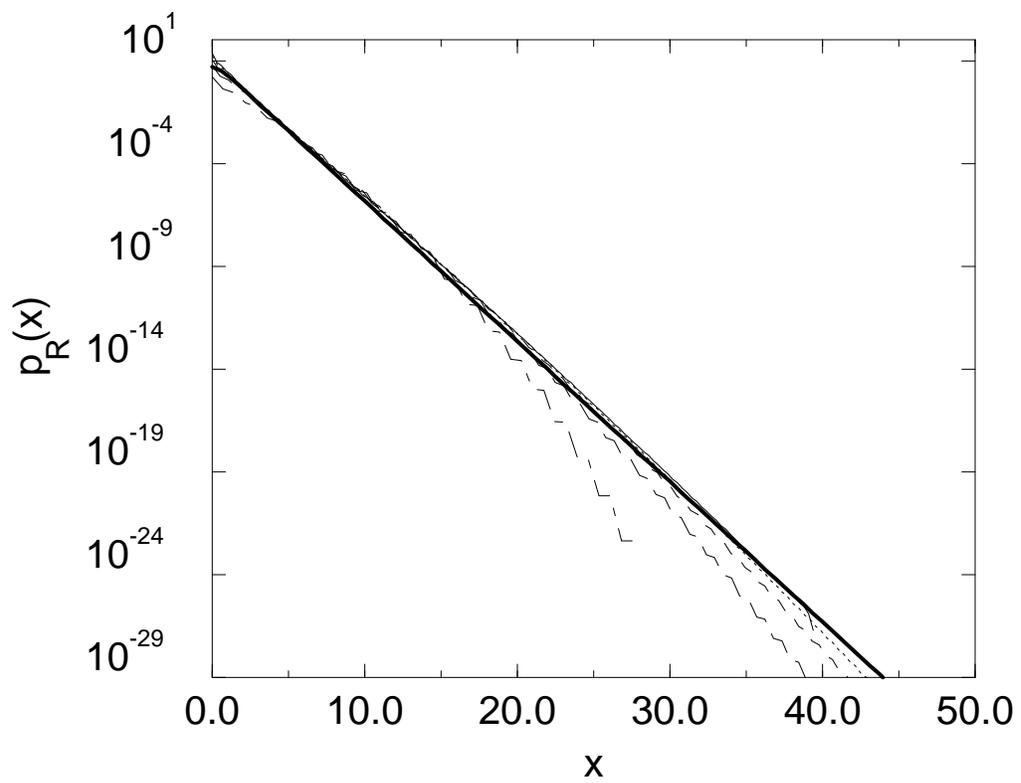
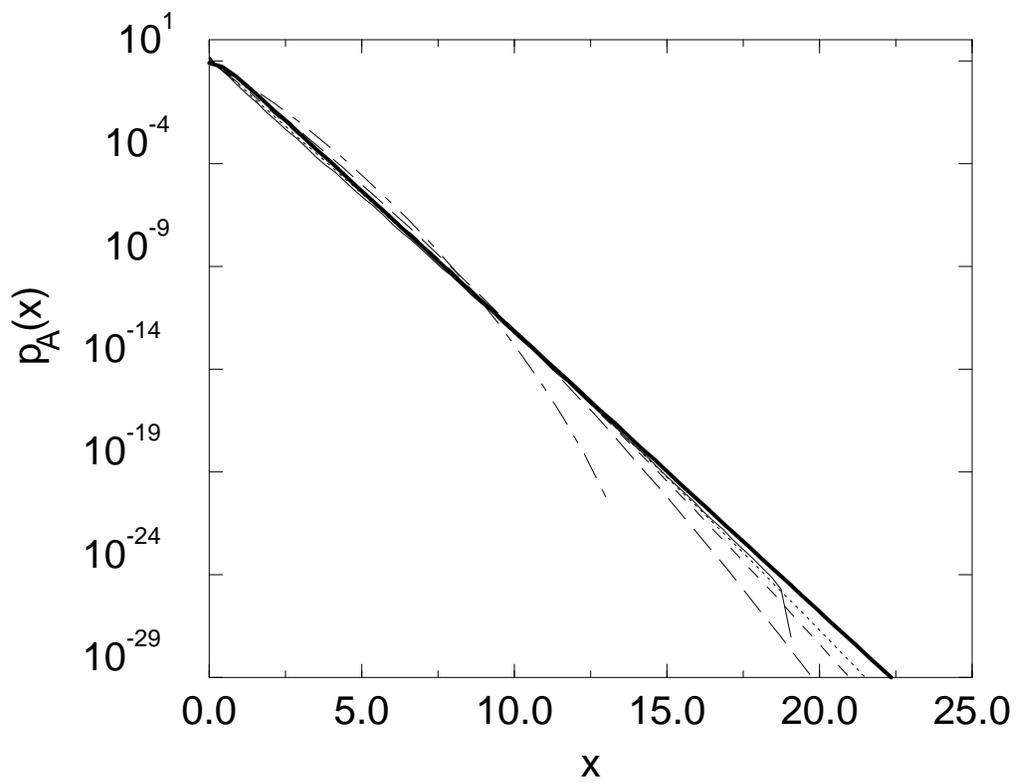
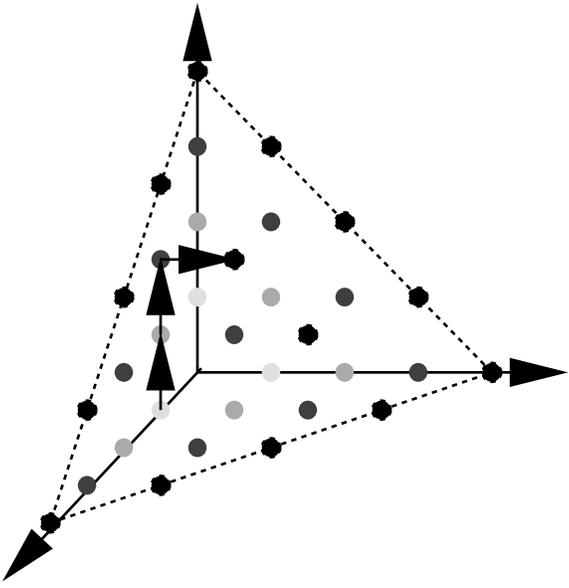
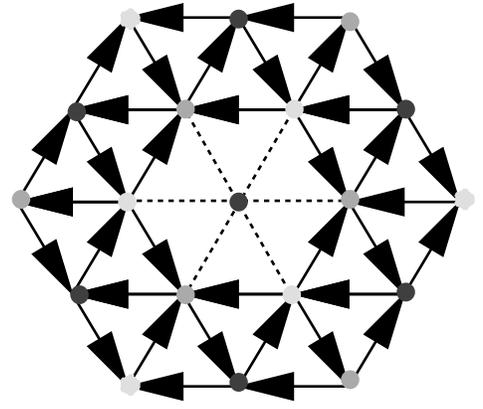
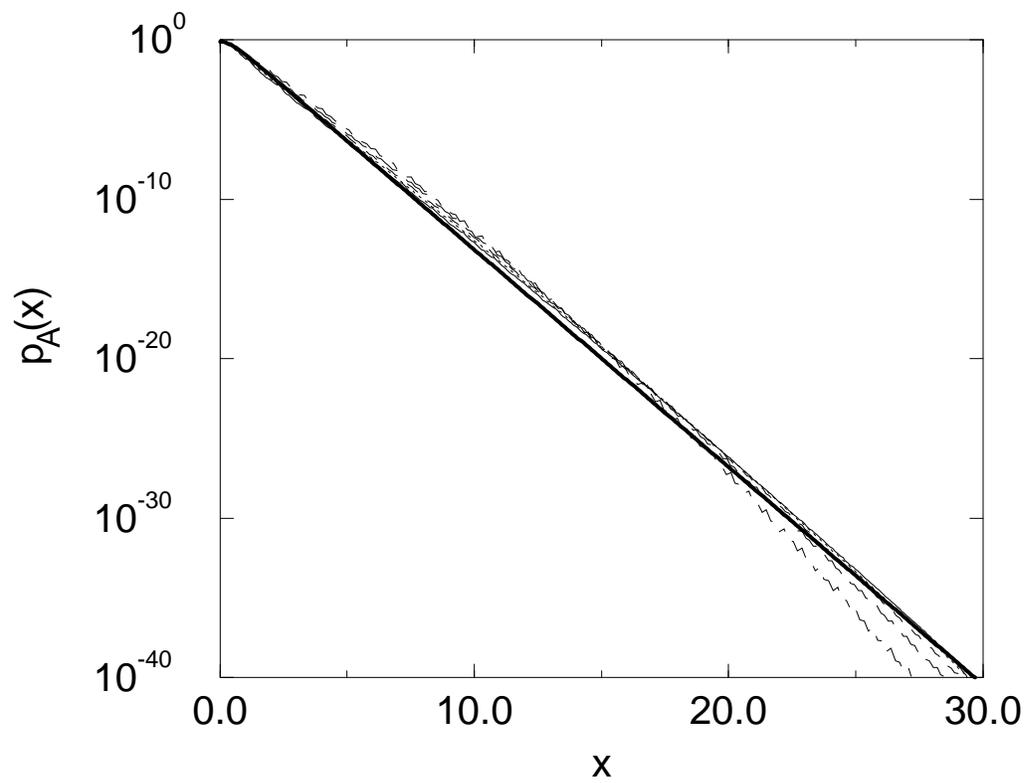
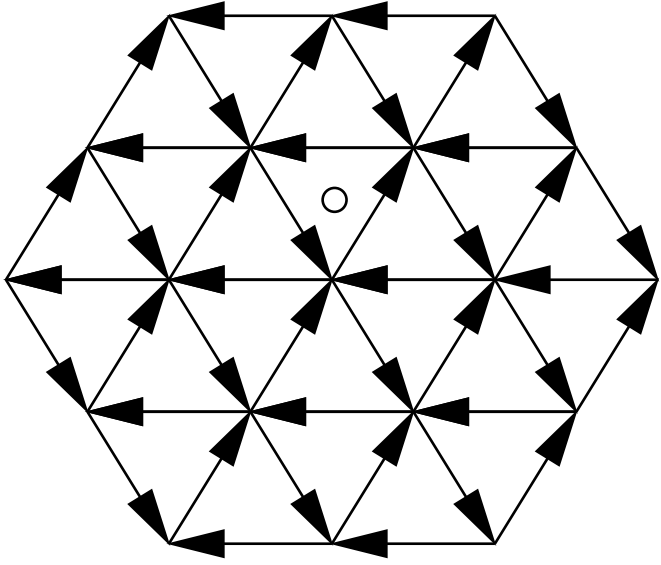
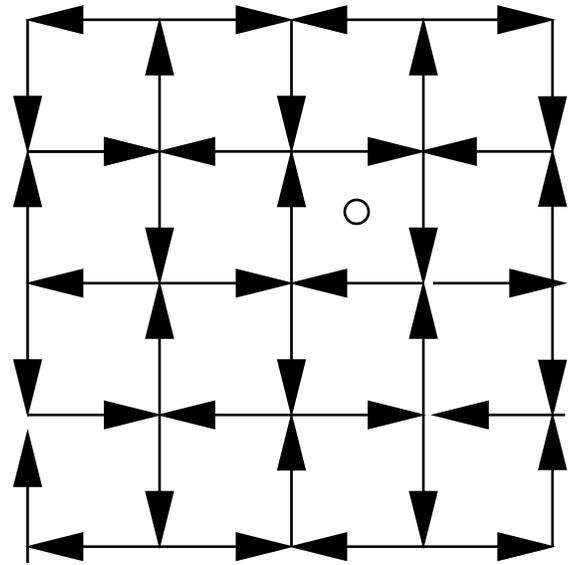
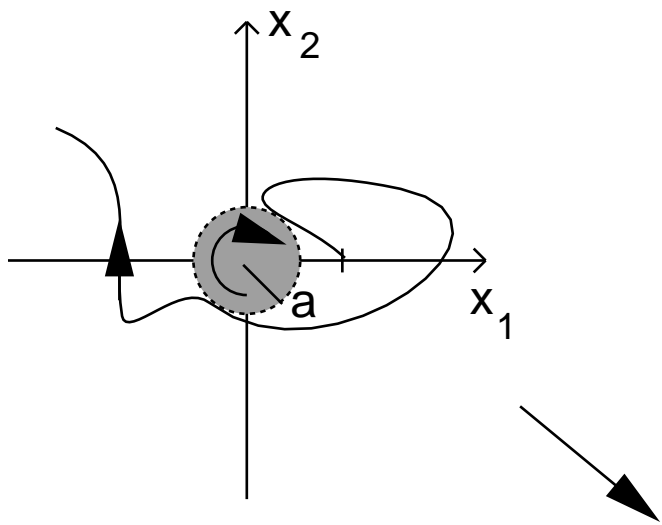
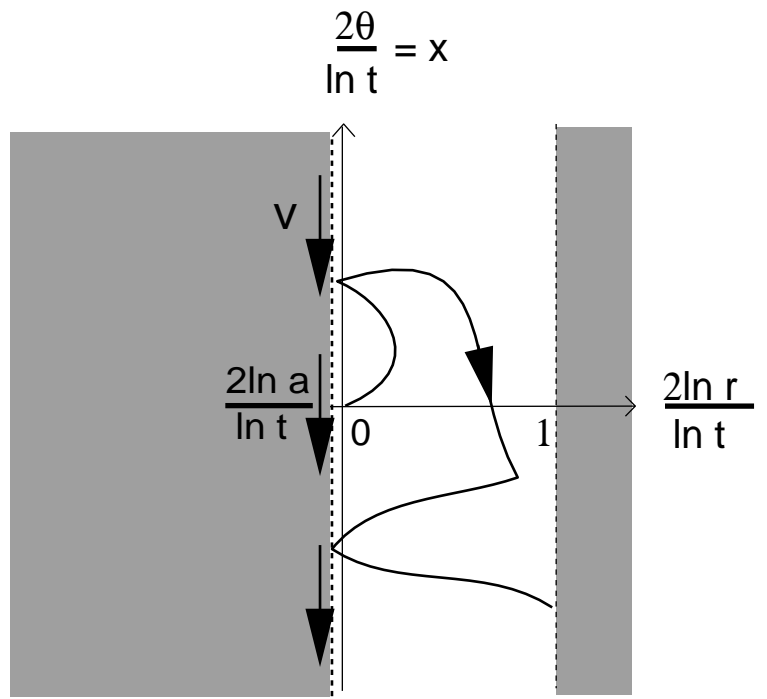
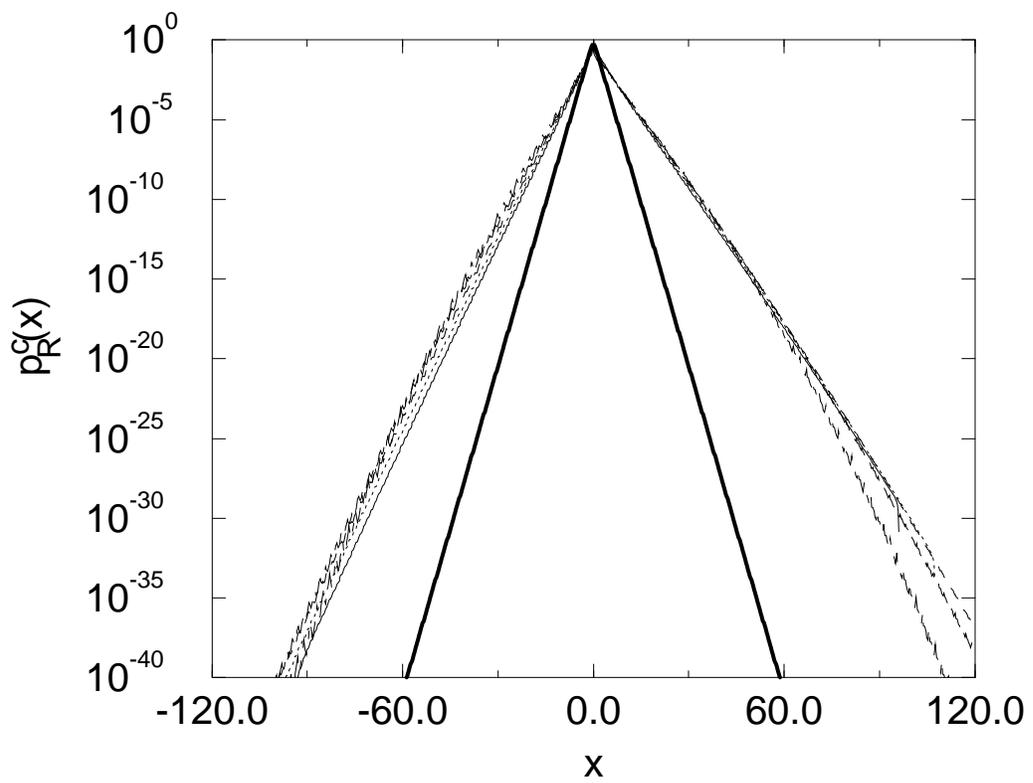
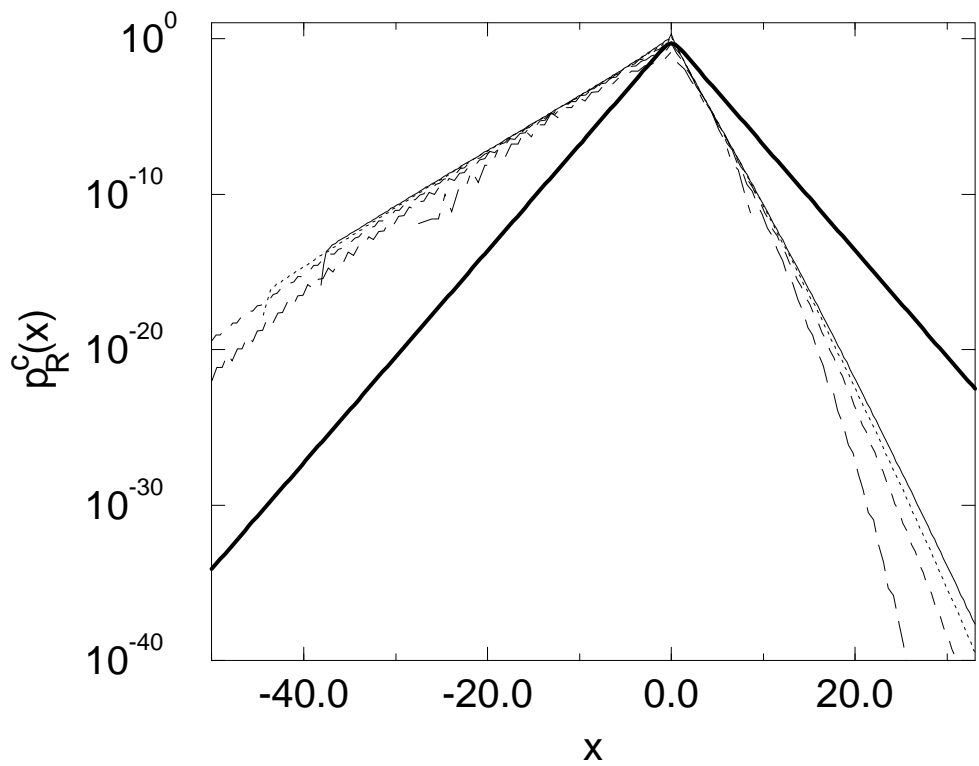
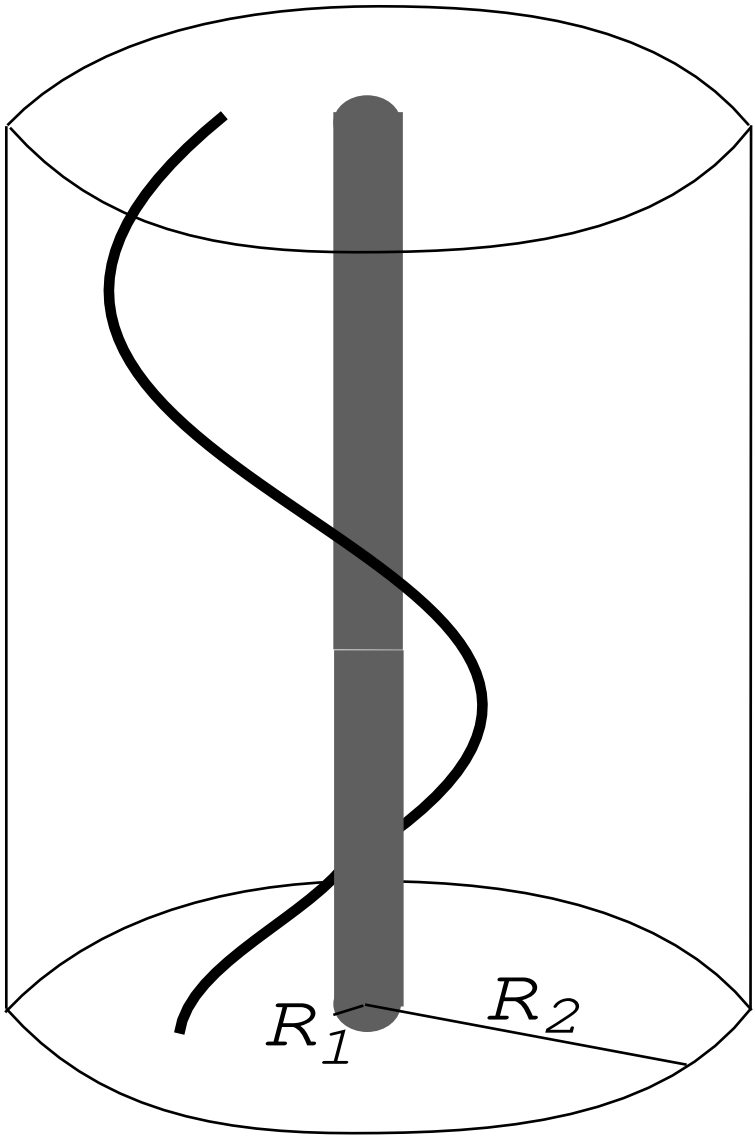
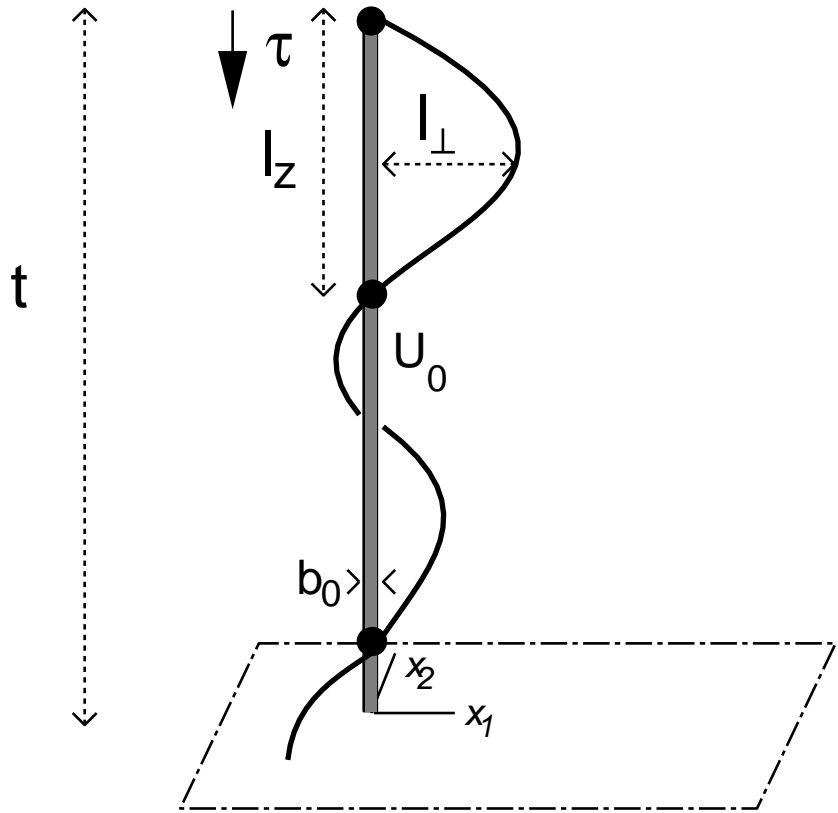
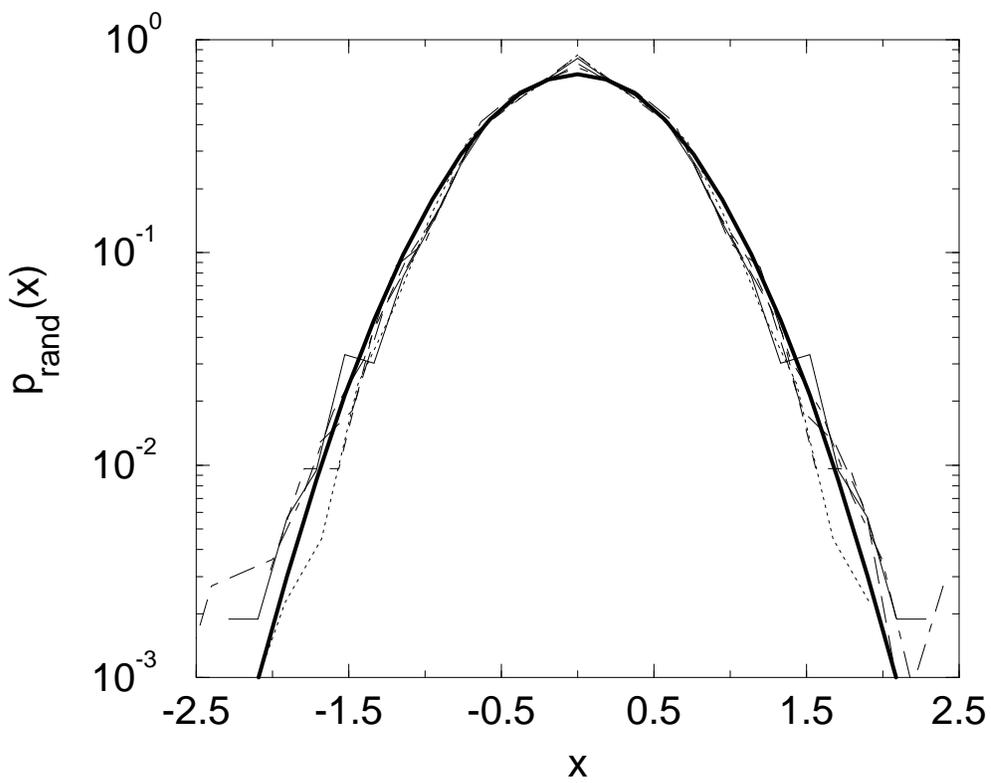
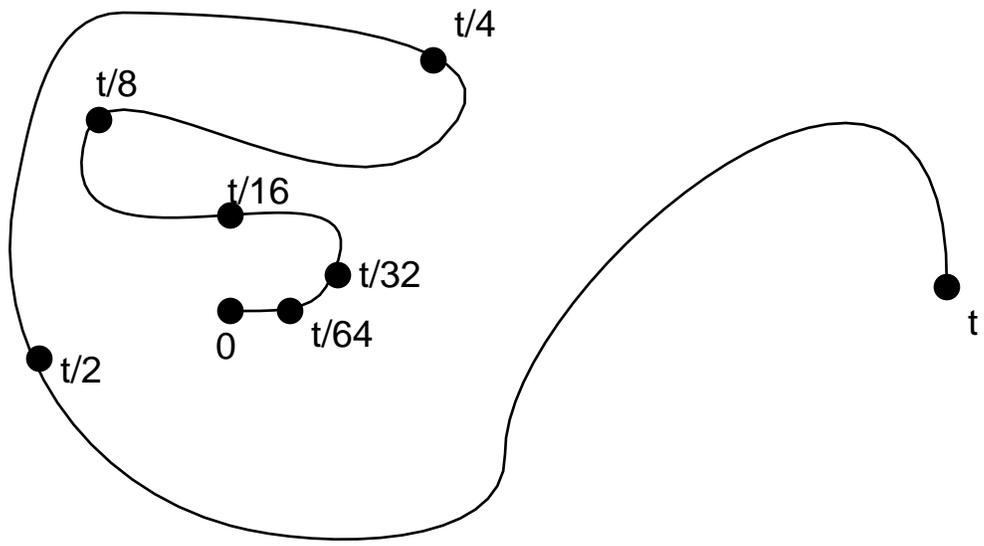
\end{document}